\documentclass[final,1p,times,twocolumn]{elsarticle}

\usepackage[dvipsnames]{xcolor}
\usepackage{amsmath}
\usepackage{amssymb}
\usepackage{graphicx}
\usepackage{bm}
\usepackage{epstopdf}
\usepackage{siunitx}
\usepackage{circuitikz}
\usepackage[mathlines]{lineno}

\usepackage{cuted}
\usepackage{capt-of}
\usepackage{microtype}
\usepackage{xurl}

\usepackage[hidelinks]{hyperref}
\graphicspath{{./}{PRL_Template/}}

\biboptions{numbers,sort&compress}
\journal{Applied Surface Science}

\begin{document}
\raggedbottom

\begin{frontmatter}

\title{Effect of Rapid Thermal Annealing on the Superconductivity of ALD-Grown TiN thin films
}

\author[aff1,aff2]{Z.~Sadirkhanov}
\author[aff3]{A.~Tuigynbek}
\author[aff2]{M.~Surendran}
\author[aff1]{A.~Jakyp}
\author[aff1]{L.~Akisheva}
\author[aff2]{A.~T.~Hammack}
\author[aff1]{A.~B.~Tikhonov\corref{cor1}}
\author[aff1,aff2]{A.~Kemelbay\corref{cor1}}

\cortext[cor1]{Corresponding authors: atikhonov@nu.edu.kz (A. Tikhonov), akemelbay@lbl.gov (A.~Kemelbay).}

\address[aff1]{Department of Physics, School of Sciences and Humanities,
Nazarbayev University, Astana, Kazakhstan}
\address[aff2]{Nanofabrication Facility, The Molecular Foundry,
Lawrence Berkeley National Laboratory, Berkeley, CA, USA}
\address[aff3]{Core Facilities, Nazarbayev University, Astana, Kazakhstan}

\begin{abstract}
Titanium nitride (TiN) is a promising superconducting material for scalable quantum technologies owing to its tunable kinetic inductance and compatibility with CMOS thin-film processing. However, the superconducting performance of atomic-layer-deposited TiN can be degraded by oxygen incorporation, off-stoichiometry, and crystallographic disorder. In this work, we investigate rapid thermal annealing (RTA) in high-purity nitrogen as a post-deposition strategy to improve polycrystalline TiN films grown by RF-biased plasma-enhanced atomic layer deposition. Grazing-incidence X-ray diffraction reveals reduced peak broadening, a smaller lattice parameter, and a stronger (111) preferred orientation with increasing annealing temperature. X-ray photoelectron spectroscopy depth profiling shows reduced oxygen-related bonding and a shift toward a Ti--N-dominated chemical environment after annealing. These structural and chemical changes result in a 25\% increase in the superconducting transition temperature and a significant narrowing of the transition width. Annealing also reduced the sheet resistance and the estimated sheet kinetic inductance, increased the residual resistance ratio, and improved the magnetic-field tolerance. Together, the structural, chemical, and transport results are consistent with reduced material disorder and improved superconducting uniformity. Because RTA is a relatively low thermal budget and is a semiconductor fabrication compatible process, it offers a practical route to improve the quality of ALD-grown TiN for superconducting applications.

\end{abstract}


\end{frontmatter}

\section{Introduction}
\label{sec:introduction}

Superconducting thin films are the building blocks of many emerging quantum technologies, including qubits, single-photon detectors, parametric amplifiers, resonators and cryogenic signal-processing circuits \cite{ref1,ref2,ref3,ref4}. As devices grow more complex, The films must simultaneously satisfy an increasingly demanding set of requirements: low microwave losses, long quasiparticle lifetimes, and stable superconducting performance well into the millikelvin regime \cite{ref5,ref6,ref7}. Meanwhile, scalable quantum hardware increasingly relies on CMOS-like multilayer fabrication flows that expose the superconducting film to high-temperature dielectric deposition, plasma etching, wafer bonding and 3D-integration steps \cite{ref8}. Each of these steps can alter the structural and chemical state of the film, so its final properties reflect not just the as-grown material but its entire processing history.

Titanium nitride is a well-studied and promising superconducting material for the fabrication of scalable quantum devices. Its composition and microstructure can be varied over a wide range, allowing the control of both the normal-state resistivity and the superconducting transition temperature, which together determine the sheet kinetic inductance \cite{ref7,ref10,ref11,ref12}.

Thus, the same material can be tuned for opposite needs: high sheet inductance for high impedance and nonlinear inductive elements and for microwave kinetic inductance detectors; or low kinetic inductance for superconducting wiring and interconnects. TiN can be easily characterized in standard sub-kelvin cryostats, having a transition temperature between 1 and 4~K \cite{ref7,ref11,ref13}.

Two main techniques are commonly used for TiN synthesis: sputtering and atomic layer deposition (ALD). Sputtering at elevated temperatures enables epitaxial growth, producing large-grained low-resistivity films with high transition temperatures. For example, Zhang et al.\ recently demonstrated that lattice-strain engineering can increase $T_{\mathrm{c}}$ in epitaxial TiN, reaching a record high $T_{\mathrm{c}}$ of 5.7~K \cite{ref14}. Although this approach deepens the understanding of intrinsic superconducting physics of the nitrides, it relies on high growth temperatures and restricts the substrate choice to lattice matched substrates, such as sapphire or MgO for TiN \cite{ref12}. In addition, sputtering, as a technique, suffers from poor step coverage, thickness gradients across large wafers and run-to-run variability, arising from target erosion and plasma drift \cite{ref15}.

In this work, we therefore focus on the ALD approach, which enables sub-nanometer thickness control, nearly ideal conformality, and excellent wafer-scale uniformity \cite{ref16}. These properties become decisive as superconducting and quantum circuits fabrication moves towards 300~mm fabrication and adopts complex layouts, such as through-silicon vias (TSVs) and chiplet based architectures \cite{ref17,ref18}. The trade-off is that ALD proceeds at comparatively lower growth temperatures, limited by precursor decomposition, and the resulting films exhibit varied performance degradation through two partly independent channels: (1) structurally, ALD films are typically nanocrystalline with a high grain boundary density \cite{ref19, Kim2003}; (2) chemically, they retain residual contamination from the precursor chemistry or the ALD reactor environment, with carbon, chlorine and oxygen being most common impurities \cite{ref10, ref16, ref20}. Xue et al.\ recently demonstrated that despite the high epitaxial quality of ALD TiN grown on sapphire at 270~$^\circ$C, the incorporation of oxygen during growth results in elevated resistivity and a reduced $T_{\mathrm{c}}$ of 2.53~K, highlighting the critical interplay between the structure and chemistry \cite{ref20}. Together these channels can contribute to defects that act as sources of disorder and scattering, which translates into raised sheet resistance, suppresses critical temperature, and broadens the superconducting transition \cite{ref21}. In addition to ALD-induced imperfections, subsequent fabrication steps used in superconducting device processing, such as elevated-temperature dielectric deposition and metal annealing, can further modify the structural and chemical state of TiN. Therefore, understanding how post-deposition annealing influences the properties of these materials is essential. When carefully controlled, annealing can serve as a useful tuning technique, allowing the material properties to be optimized according to the specific requirements of the application.

In this work, we show that rapid thermal annealing (RTA) of ALD-grown TiN in a high-purity nitrogen environment improves its superconducting properties. We study the evolution of chemical, crystallographic, electrical, and superconducting properties of thin films as a function of annealing temperature between 400 and 900~$^\circ$C. The results show that RTA reduces oxygen-related bonding, improves crystallographic ordering, concurrently increases $T_{\mathrm{c}}$ by 25\% from 2.4~K to 3.0~K, sharpens the superconducting transition, and lowers the sheet resistance. In contrast to approaches based on epitaxial strain engineering, the present method addresses the oxygen-related chemical disorder characteristic of ALD-grown TiN using a short and CMOS compatible post-deposition treatment. RTA can therefore be introduced immediately after TiN deposition or at thermally compatible intermediate stages of a multilevel fabrication process, providing a practical route for improving the performance of ALD-based superconducting films.

\section{Methods}
\label{sec:methods}

Titanium nitride (TiN) thin films were synthesized using an Oxford Instruments FlexAL plasma-enhanced atomic layer deposition (PEALD) system. Deposition was performed on 4-inch intrinsic Si(100) wafers at $300\,^{\circ}\mathrm{C}$. Tetrakis(dimethylamido)titanium (TDMAT, 99.99\%, Strem Chemicals), heated to $75\,^{\circ}\mathrm{C}$, was used as the Ti precursor, while an N$_2$/H$_2$ plasma was used as the co-reactant. Each ALD cycle consisted of a 500~ms TDMAT pulse followed by a 2~s Ar purge. The reaction was completed using a 2~s inductively coupled plasma (ICP) step, followed by a 5~s RF-bias step and a 2~s Ar purge. The ICP and substrate RF-bias powers were set to 100 and 40~W, respectively. The growth rate was monitored using a Woollam iSE spectroscopic ellipsometer, yielding a growth-per-cycle rate of approximately $0.59~\text{\AA}/\mathrm{cycle}$. A total of 1020 cycles was used to obtain 60~nm TiN films, with the film thickness additionally confirmed by cross-sectional scanning electron microscopy (SEM).

Rapid thermal annealing (RTA) of the TiN thin films was performed using an ULVAC MILA-5000 system. Annealing was carried out at atmospheric pressure under a continuous flow of high-purity N$_2$ (99.999\%) through a quartz tube at a flow rate of 2~L/min. The N$_2$ flow maintained a slight positive pressure within the tube to minimize diffusion of ambient air into the annealing environment. The thermal treatment consisted of two steps. First, the samples were conditioned at $110\,^{\circ}\mathrm{C}$ for 10~min to reduce residual moisture and air within the system. This was followed by RTA at temperatures ranging from 400 to $900\,^{\circ}\mathrm{C}$ for 30~s, with an average ramp rate of approximately $50\,^{\circ}\mathrm{C\,s^{-1}}$. The conditioning step was introduced after preliminary experiments showed improved reproducibility of the annealing results following its implementation.

Superconducting properties were measured using a Quantum Design PPMS DynaCool system with a base temperature of 1.8~K and a 14~T superconducting magnet. Electrical measurements were performed using the resistivity module in a four-point configuration with an excitation current of $1~\mu\mathrm{A}$. Temperature-dependent resistance measurements were performed at a sweep rate of $\pm 0.5~\mathrm{K\,min^{-1}}$, while magnetic-field sweeps were performed at $\pm 100~\mathrm{Oe\,s^{-1}}$. To investigate the effect of annealing on thinner TiN films with superconducting transition temperatures below 1~K, additional temperature-dependent resistance measurements were performed on 20~nm TiN films using a High Precision Devices Model 106 Shasta pulse-tube-driven adiabatic demagnetization refrigerator (ADR) with a base temperature of approximately 100~mK.

X-ray photoelectron spectroscopy (XPS) measurements were performed using a Thermo Scientific NEXSA system equipped with an Al K$\alpha$ X-ray source ($h\nu = 1486$~eV). Depth profiling was performed using a monatomic Ar$^{+}$ ion gun operated at 2~keV to remove the native surface oxide and probe the underlying TiN film. Ten sequential sputtering cycles, each 30~s in duration, were performed, with XPS spectra acquired between successive cycles. The evolution of the elemental composition during sputter depth profiling is shown in Fig.~S2 of the Supporting Information. After the initial sputtering stage, the Ti, N, O, and C concentrations approached comparatively stable values. The spectra discussed in the main text were therefore acquired after the tenth sputtering cycle to compare the underlying TiN films after removal of the dominant surface-oxide contribution. Identical sputtering and acquisition conditions were used for all samples to enable direct comparison across the annealing series. No systematic peak shifts or additional peak broadening were observed during the later stages of depth profiling.

Crystallographic properties were examined using a Rigaku SmartLab X-ray diffraction (XRD) system with Cu K$\alpha$ radiation ($\lambda = 1.5406$~\text{\AA}). Measurements were performed in grazing-incidence X-ray diffraction (GI-XRD) geometry at a fixed incidence angle of $1^{\circ}$ to reduce the substrate contribution and enhance sensitivity to diffraction from the TiN thin film.

\clearpage
\section{Results and Discussion}
\label{sec:results}

\subsection{Superconductive and electrical properties}
\label{sec:Tc}

\begin{center}
\begin{minipage}{\textwidth}
\centering
\includegraphics[width=\linewidth]{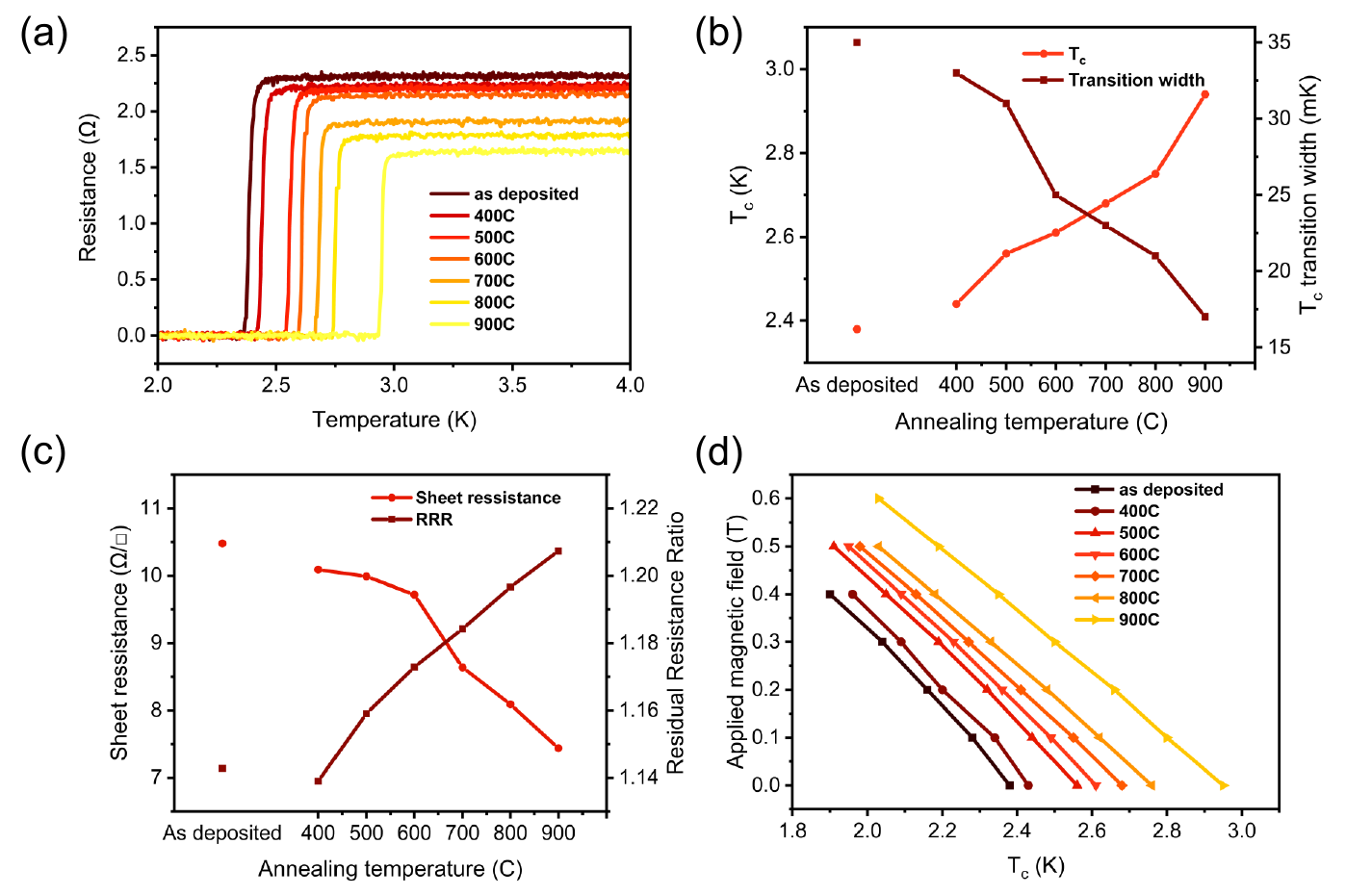}

\captionof{figure}{Evolution of the electrical and superconducting
properties of TiN thin films with RTA temperature.
(a) Superconducting transitions of TiN films annealed at
different temperatures;
(b) $T_{\mathrm{c}}$ and superconducting transition width
as functions of annealing temperature;
(c) sheet resistance and residual resistance ratio (RRR)
as functions of annealing temperature; and
(d) magnetic-field dependence of $T_{\mathrm{c}}$
for differently annealed samples.}
\label{fig:electrical}
\end{minipage}
\end{center}

Figure~\ref{fig:electrical}(a) shows the superconducting transition measurements of as-deposited and annealed 60~nm-thick TiN films. The temperature-dependent resistance curves show that the superconducting transition temperature increases monotonically with annealing temperature, while the transition itself becomes markedly narrower. Each transition curve was fitted with a Boltzmann sigmoid function, from which $T_{\mathrm{c}}$ is taken as the midpoint of the fit, and the transition width, $\Delta T$, is defined as the temperature interval over which the resistance increases from 10\% to 90\% of the normal-state resistance ($R$). Figure~\ref{fig:electrical}(b) plots both quantities as functions of annealing temperature. $T_{\mathrm{c}}$ increases by 25\%, from 2.4 to 3.0~K, where it begins to saturate, while $\Delta T$ narrows from 35 to 15~mK. Both trends are consistent with a reduction in spatial disorder in the polycrystalline superconducting film sampled along the current path~\cite{ref21,ref22}. The effect is seen most clearly in a 20~nm TiN film annealed at 600~$^{\circ}$C (Fig.~S4 in the Supporting Information), where the as-deposited film exhibits a two-step transition, suggesting regions with different $T_{\mathrm{c}}$ values. RTA eliminates the lower-temperature transition step and leaves a narrower transition at a higher temperature.

Figure~\ref{fig:electrical}(c) shows the residual resistance ratio (RRR) and the normal-state sheet resistance as functions of annealing temperature. The sheet resistance decreases systematically from approximately $10.5~\Omega/\square$ to $7.1~\Omega/\square$, while the RRR increases from 1.140 to 1.215. All films exhibit $\mathrm{RRR}>1$, and the increase in RRR indicates a larger relative resistance change between room temperature and the low-temperature normal state. Taken together, the systematic increase in $T_{\mathrm{c}}$ and RRR, the narrowing of the superconducting transition, and the decrease in sheet resistance are consistent with reduced disorder-related scattering. Point defects and grain boundaries are possible scattering sources in nanocrystalline TiN, but their individual contributions cannot be distinguished from these transport measurements alone~\cite{ref12}. The structural and compositional evolution, discussed below, is also consistent with previous reports linking thermal treatment and oxygen incorporation to changes in TiN electrical properties~\cite{ref21,ref23}.

\subsection{Crystallographic properties}
\label{sec:xrd}

\begin{figure*}[!t]
\centering
\includegraphics[width=\textwidth]{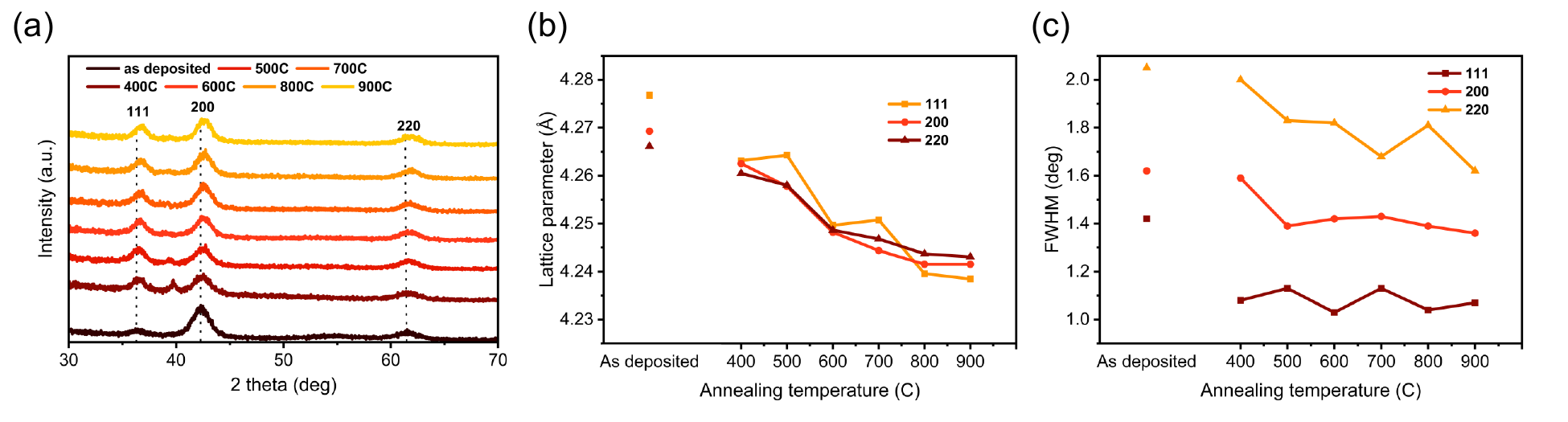}
\caption{Structural evolution of TiN thin films
with annealing temperature.
(a) Grazing-incidence X-ray diffraction (GI-XRD) patterns
of TiN samples;
(b) calculated lattice parameter as a function of
annealing temperature; and
(c) full width at half maximum (FWHM) as a function of
annealing temperature.}
\label{fig:xrd}
\end{figure*}

Figure~\ref{fig:xrd}(a) shows grazing-incidence XRD patterns of the as-deposited and annealed films, measured at an incidence angle of 1$^{\circ}$. All diffraction peaks are indexed to rocksalt TiN, and no secondary crystalline phases are observed, indicating that no detectable crystalline titanium oxide or silicide forms upon annealing at temperatures up to 900 °C. To quantify the evolution of the diffraction patterns, each peak was fitted with a Gaussian profile to extract its position and full width at half maximum (FWHM).
The apparent lattice parameters obtained from the 111, 200, and 220 reflections decrease from 4.277, 4.269, and 4.266~\AA{} in the as-deposited film to 4.231, 4.239, and 4.239~\AA{} after annealing at 900~$^{\circ}$C, approaching the reported the reported relaxed lattice parameter of stoichiometric TiN (4.240~\AA{})~\cite{ref24,ref25}. The as-deposited film therefore exhibits an apparent lattice expansion of up to approximately 0.9\% relative to this reference value. Additionally, the lattice parameters obtained from the different reflections are not identical, and their spread naroows after annealing. These observations likely reflect a combination of residual strain, composition-related lattice distortions, and structural defects. Composition-dependent changes in TiN lattice spacing and the presence of stacking faults associated with grain boundaries have been reported previously ~\cite{ref26}. Stacking and twin faults can also induce reflection-dependent changes in diffraction-peak positions and profiles, providing another possible contribution to the differences between the apparent lattice parameters~\cite{ref27}.

The FWHM values generally decrease with increasing annealing temperature, although some non-monotonic variations are observed. This narrowing is attributed to a combination of crystallite growth, microstrain relaxation, and the reduction of structural defects~\cite{ref27,ref28}. To examine the origin of the broadening, a qualitative Williamson--Hall analysis was performed (Fig.~S5, Supporting Information). In the GI-XRD geometry, each reflection samples a different subset of grains, namely those whose diffracting planes are inclined by the corresponding angle $\psi$. The Williamson--Hall plot therefore compares grain populations of different orientations, but it still provides a useful indication of reflection-dependent broadening. The as-deposited and 400~$^{\circ}$C films exhibit strongly reflection-dependent broadening, most notably an excess broadening of the (200) reflection. Because TiN is elastically stiffest along $\langle100\rangle$, anisotropic strain broadening alone would be expected to make the (200) reflection narrower rather than broader. The excess (200) broadening is therefore more consistent with stacking faults, which in face-centered cubic (fcc) lattices can broaden the (200) reflection more strongly than the (111) reflection. Upon annealing at 500~$^{\circ}$C and above, this excess broadening disappears and the data follow an approximately linear trend, consistent with a reduction in planar defects and microstrain. Concurrently, the (111) reflection shows the most pronounced narrowing and an increase in relative intensity. Since the probing depth is primarily determined by the fixed incidence angle and is approximately comparable across the measured reflections, this change may reflect an increased fraction of grains whose $\{111\}$ planes are inclined by approximately $17^{\circ}$ to the sample surface, suggesting an evolution in crystallographic texture. Confirming a (111) preferred orientation would require additional symmetric $\theta$--$2\theta$ scans or pole-figure measurements. Taken together, the evolution of the apparent lattice parameters, peak broadening, and crystallographic texture, particularly the disappearance of the possible stacking-fault signature near 500~$^{\circ}$C, coincides with the increase in superconducting transition temperature and the narrowing of the transition width observed in the transport measurements.
\begin{figure*}[!t]
    \centering
    \includegraphics[
        width=\textwidth,
        height=0.78\textheight,
        keepaspectratio
    ]{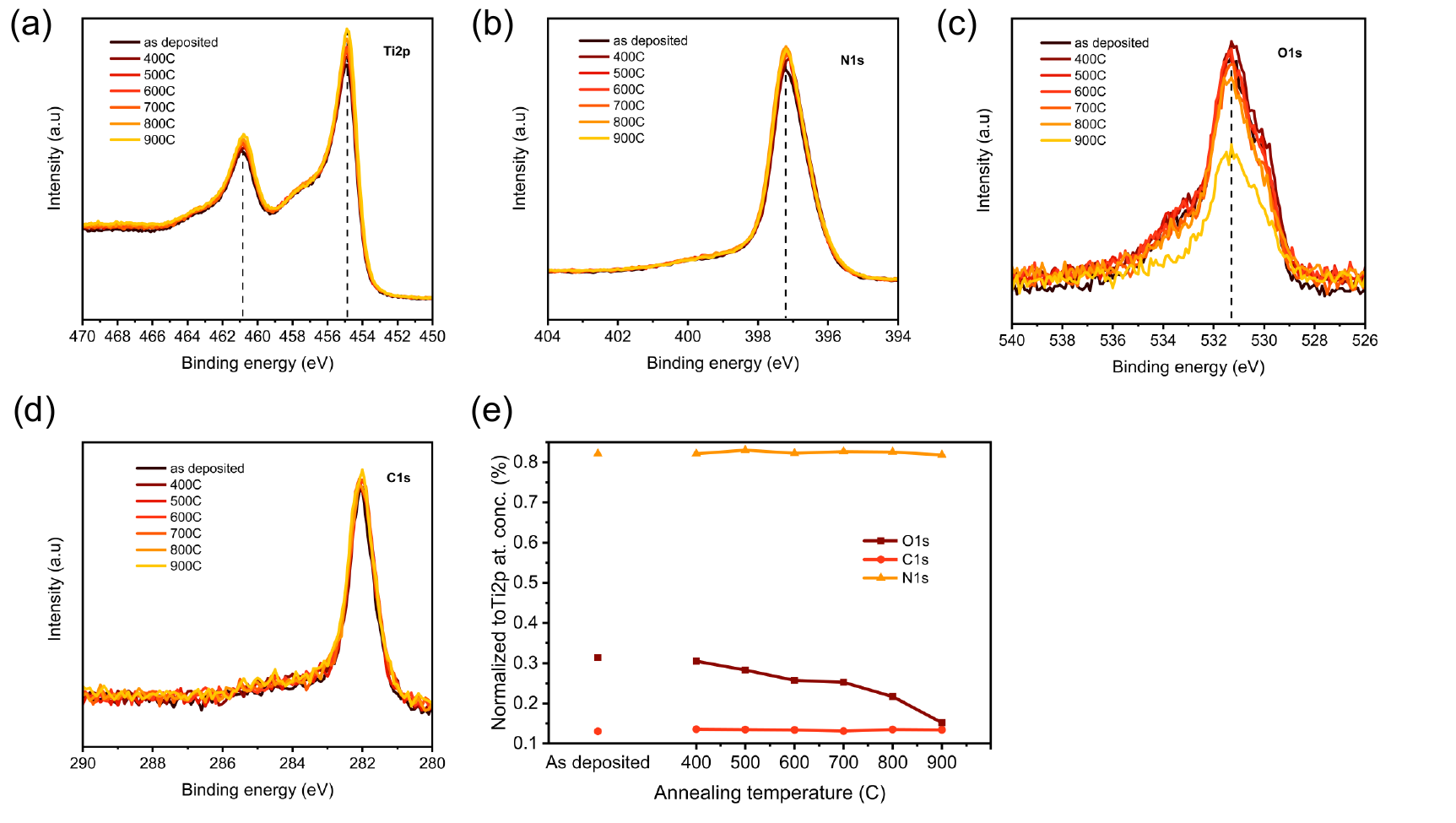}
   \caption{XPS analysis of the TiN thin films as a function of annealing temperature:
(a) Ti 2p, (b) N 1s, (c) O 1s, and (d) C 1s core-level spectra;
(e) elemental atomic ratios (O/Ti, C/Ti, and N/Ti) as a function of annealing temperature.}
    \label{fig:figure3}
\end{figure*}

\subsection{XPS analysis}
\label{sec:xps}

 X-ray photoelectron spectroscopy (XPS) was used to investigate the stoichiometry of TiN thin films. To probe the film beneath the native surface oxide, each sample was sequentially sputter-etched and a spectrum was recorded. The spectra presented here were acquired once a compositional steady state had been reached, and no peak shifts were observed between successive sputter cycles. The Ti~2p envelope can contain overlapping contributions from TiN, titanium oxides, and oxynitrides, as well as intrinsic shake-up and energy-loss features~\cite{ref29,ref30}. Because these contributions are difficult to separate reliably, the Ti 2p spectra were analyzed qualitatively. Figure~\ref{fig:figure3}(a) shows the Ti~2p region after removal of the surface oxide, with two main peaks at 455 and 461~eV, indicated by dashed lines and corresponding to the 2p$_{3/2}$ and Ti 2p$_{1/2}$ doublet of TiN. The overall envelope shape does not change substantially with annealing temperature, apart from a slight increase in the intensity of the TiN component, consistent with the evolution of the N 1s peak (Figure~\ref{fig:figure3}(b)). Oxygen may have been incorporated during the prolonged ALD growth or upon subsequent air exposure (Figure~\ref{fig:figure3}(c)), whereas carbon most likely originates from the methyl-rich ALD precursor (Figure~\ref{fig:figure3}(d)). RTA has no observable effect on the carbon concentration, which remains at approximately 6~at.\%.

In contrast, the O~1s peak intensity shows the largest change, decreasing significantly with increasing annealing temperature (Figure~\ref{fig:figure3}(c)). The absolute oxygen content may be overestimated because residual gases in the XPS chamber, with a base pressure of $5 \times 10^{-8}$~Torr, can re-adsorb onto the highly reactive, freshly sputtered surface ~\cite{ref31}. Since all samples were measured under identical conditions, however, this contribution should affect them similarly, and the relative trend remains meaningful. Figure~\ref{fig:figure3}(e) shows the  O/Ti, C/Ti, and N/Ti ratios, revealing that annealing has the largest effect on the measured oxygen-to-titanium ratio, reducing it by approximately 50\%. We propose two mechanisms that may contribute to this reduction. First, annealing at higher temperatures may promote the removal of interstitial or weakly bound oxygen from the film, leading to a more stoichiometric TiN composition~\cite{ref32}. Second, annealing may increase the grain size, decreasing the grain-boundary area available for oxidation upon subsequent air exposure. The decrease in the measured oxygen content is accompanied by an increase in RRR [Figure~\ref{fig:electrical}(c)], consistent with a shift toward more metallic electrical behavior. These trends are in line with previous studies of oxygen incorporation and oxidation in TiN~\cite{ref20,ref33}.

Annealing therefore induces both structural and chemical changes. A closer examination of the evolution of $T_{\mathrm{c}}$ [Figure~\ref{fig:electrical}(b)] shows that these changes occur over overlapping temperature ranges. However, the two pronounced increases in $T_{\mathrm{c}}$, between 400 and 500~$^{\circ}$C and between 800 and 900~$^{\circ}$C, suggest that different mechanisms dominate at different annealing temperatures.

We attribute that the first significant increase in $T_{\mathrm{c}}$, between 400 and 500 °C, primarily to a reduction in crystallographic disorder. In the qualitative Williamson–Hall analysis (Fig. S5, Supporting Information), this temperature window coincides with the abrupt disappearance of the excess broadening of the (200) reflection. This behavior indicates structural ordering and possibly a reduction in planar defects, although a three-reflection analysis cannot uniquely identify the defects involved. Such structural reorganization lengthens the electron mean free path, which may weaken the disorder-enhanced Coulomb repulsion known to suppress Tc in strongly disordered TiN ~\cite{ref_kruhlov}. Improved structural homogeneity may also reduce spatial variations in the superconducting properties, consistent with the concurrent narrowing of the transition.

The second pronounced increase in $T_{\mathrm{c}}$, between 800 and 900~$^{\circ}$C, coincides with a further change in chemical composition. The normalized XPS results [Figure~\ref{fig:figure3}(e)] show that the O/Ti ratio decreases gradually at lower annealing temperatures and then drops more sharply between 800 and 900~$^{\circ}$C. No additional crystalline phases are detected by GI-XRD, indicating that this change is not associated with the formation of crystalline oxide or silicide phases, although amorphous grain-boundary oxides would not be detected. A possible explanation is that higher temperatures facilitate the removal or redistribution of oxygen-related species. The present measurements cannot distinguish the removal of interstitial oxygen from changes in local oxide or oxynitride environments. Oxygen incorporated into the TiN lattice is known to lower $T_{\mathrm{c}}$ by modifying the electronic structure \cite{ref10}, and oxygen at grain boundaries can increase disorder and inhomogeneity; a reduction in either would contribute to the improved superconducting properties. Overall, the results are consistent with a two-stage response to RTA: structural reorganization dominates at intermediate temperatures, followed by a more pronounced change in oxygen-related chemistry at higher temperatures. Because these processes overlap, their individual contributions cannot be isolated from the present data.

\section{Conclusion}
\label{sec:conclusion}

Rapid thermal annealing in high-purity nitrogen is an effective post-deposition treatment for improving the superconducting and electrical properties of ALD-grown TiN thin films. For the 60~nm films, annealing increased $T_{\mathrm{c}}$ by 25\%, from 2.4 to 3.0~K, narrowed the superconducting transition width from 35 to 17~mK, and reduced the normal-state sheet resistance by approximately 32\%, from 10.5 to 7.1~$\Omega/\square$. The residual resistance ratio increased from 1.140 to 1.215, while the magnetic-field-dependent transitions shifted toward higher temperatures. Similar improvements were observed in a 20~nm film, where annealing transformed a broad, two-step superconducting transition into a narrower transition at a higher temperature.

GI-XRD shows that the rocksalt TiN structure is retained, with no additional crystalline phases detected within the sensitivity of the measurements. RTA shifts the apparent lattice parameters toward the reference value for stoichiometric TiN and reduces diffraction-peak broadening, while the increased relative intensity of the (111) reflection suggests texture evolution. Qualitative Williamson--Hall analysis further indicates reduced reflection-dependent broadening following higher-temperature annealing. These observations are consistent with structural reorganization, although the three-reflection analysis cannot uniquely separate crystallite-size, strain, and planar-defect contributions. XPS depth profiling reveals an approximately 50\% decrease in the measured oxygen-to-titanium ratio, accompanied by slightly increased Ti--N and N~1s contributions, while the carbon concentration remains near 6~at.\%. These changes indicate a shift toward a more Ti--N-dominated chemical environment.

Together, the transport, structural, and chemical measurements establish a correlation between reduced oxygen content, annealing-induced structural evolution, and improved superconducting behavior. Reduced chemical disorder and structural reorganization are plausible contributors to the higher $T_{\mathrm{c}}$, sharper transition, and lower sheet resistance. However, their individual contributions cannot be isolated because composition, lattice spacing, crystallographic ordering, and texture evolve concurrently. The short RTA treatment therefore offers a practical route for tailoring ALD-grown TiN after deposition for superconducting resonators, detectors, interconnects, and quantum circuits.

\section*{Acknowledgements}
\label{sec:acknowledgements}
This research is funded by the Science Committee of the Ministry of Science and Higher Education of the Republic of Kazakhstan (Grant No. AP32731267).

Work at the Molecular Foundry was supported by the Office of Science, Office of Basic Energy Sciences, of the U.S. Department of Energy under Contract No. DE-AC02-05CH11231, under user proposal number MFP-9551.

\section*{CRediT authorship contribution statement}
\label{sec:credit}
\textbf{Z. Sadirkhanov:} Conceptualization, Methodology, Investigation, Formal analysis, Data curation, Visualization, Writing – original draft, Writing – review \& editing.
\textbf{A. Tuigynbek:} Methodology, Investigation, Writing – review \& editing.
\textbf{M. Surendran:} Conceptualization, Methodology, Investigation, Writing – original draft, Writing – review \& editing.
\textbf{A. Zhakyp:} Investigation, Writing – review \& editing.
\textbf{L. Akisheva:} Investigation, Writing – review \& editing.
\textbf{A. T. Hammack:} Conceptualization, Writing – review \& editing.
\textbf{A. Tikhonov:} Conceptualization, Writing – review \& editing, Funding acquisition, Project administration.
\textbf{A. Kemelbay:} Conceptualization, Methodology, Supervision, Investigation, Writing – review \& editing.

\section*{Data availability}
\label{sec:data}
Data will be made available on request.

\section*{Declaration of generative AI use}
\label{sec:AI}
During the preparation of this work the authors used LLM to prepare the graphical abstract, and for general brainstorming. After using this tool, the authors reviewed and edited the content as needed and take full responsibility for the content of the published article.

\end{document}